# Formation of chlorinated breakdown products during degradation of sunscreen agent, 2-ethylhexyl-4-methoxycinnamate in the presence of sodium hypochlorite

Alicja Gackowska, Maciej Przybyłek, Waldemar Studziński & Jerzy Gaca

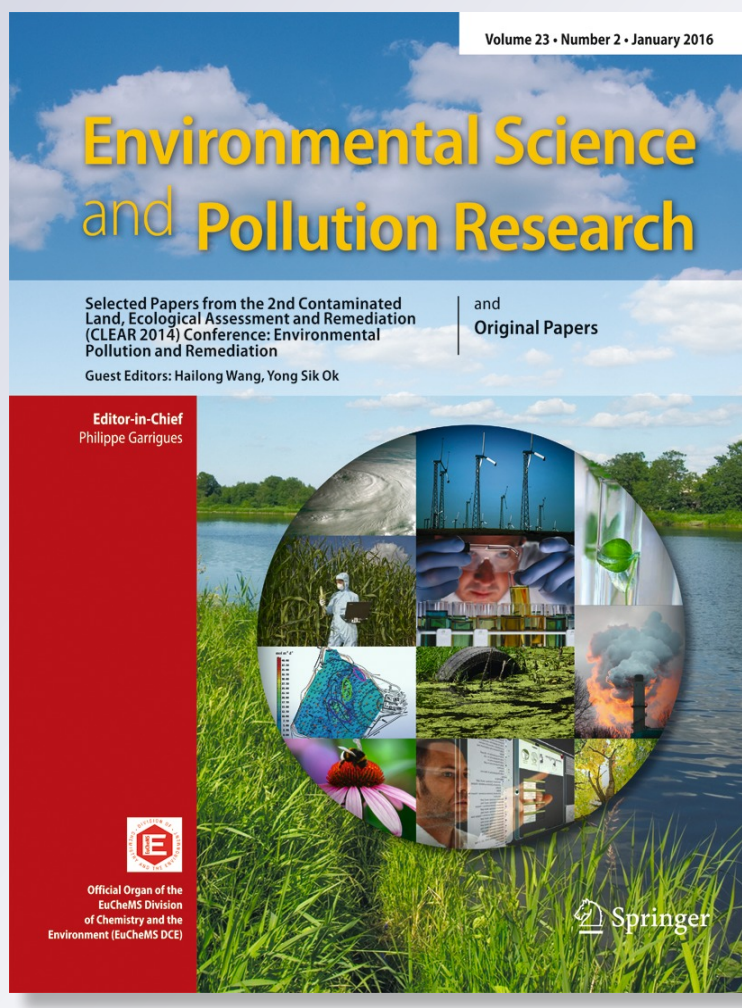

# Formation of chlorinated breakdown products during degradation of sunscreen agent, 2-ethylhexyl-4-methoxycinnamate in the presence of sodium hypochlorite

Alicja Gackowska[1] · Maciej Przybyłek[2] · Waldemar Studziński[1] · Jerzy Gaca[1]

**Abstract** In this study, a new degradation path of sunscreen active ingredient, 2-ethylhexyl-4-methoxycinnamate (EHMC) and 4-methoxycinnamic acid (MCA) in the presence of sodium hypochlorite (NaOCl), was discussed. The reaction products were detected using gas chromatography–mass spectrometry (GC-MS). Since HOCl treatment leads to more polar products than EHMC, application of polar extracting agents, dichloromethane and ethyl acetate/n-hexane mixture, gave better results in terms of chlorinated breakdown products identification than n-hexane. Reaction of EHMC with HOCl lead to the formation of C=C bridge cleavage products such as 2-ethylhexyl chloroacetate, 1-chloro-4-methoxybenzene, 1,3-dichloro-2-methoxybenzene, and 3-chloro-4-methoxybenzaldehyde. High reactivity of C=C bond attached to benzene ring is also characteristic for MCA, since it can be converted in the presence of HOCl to 2,4-dichlorophenole, 2,6-dichloro-1,4-benzoquinone, 1,3-dichloro-2-methoxybenzene, 1,2,4-trichloro-3-methoxybenzene, 2,4,6-trichlorophenole, and 3,5-dichloro-2-hydroxyacetophenone. Surprisingly, in case of EHMC/HOCl/UV, much less breakdown products were formed compared to non-UV radiation treatment. In order to describe the nature of EHMC and MCA degradation, local reactivity analysis based on the density functional theory (DFT) was performed. Fukui function values showed that electrophilic attack of HOCl to the C=C bridge in EHMC and MCA is highly favorable (even more preferable than phenyl ring chlorination). This suggests that HOCl electrophilic addition is probably the initial step of EHMC degradation.

## Introduction

Sodium hypochlorite is one of the most commonly used disinfectants in water treatment. Under the aqueous conditions it can be hydrolyzed to hypochlorous acid (HOCl). Since HOCl reveal a high chlorinating and oxidizing activity, it is often classified both to the reactive oxygen and chlorine species. The pollutants which are formed via dissolved organic matter conversion in the presence of chlorine disinfectants are called halogenated disinfection by-products (HDBPs). Generally speaking, this class of pollutants comprises trihalomethanes (THMs), haloacetic acids, haloacetonitriles (HANs), and haloketones (HKs) (Boorman 1999). HDBPs can be formed through the transformation of naturally occurring compounds like humic and fulvic acids (Ristoiu et al. 2009). However, many studies have been also focused on purely anthropogenic HDBPs (Negreira et al. 2008; Nakajima et al. 2009; Santos et al. 2013; Solakyildirim et al. 2014; Bulloch et al. 2015) which can be formed through pharmaceuticals and personal care product (PPCP) reactions with disinfectants. The use of PPCPs has dramatically increased since the late 1990s. It is estimated that between 1999 and 2009, the annual PPCP consumption in the USA has grown from 2 billion to 3.9 billion (Tong et al. 2011). Since the influence of PPCPs on the ecosystems has not been fully studied, they are often classified as emerging pollutants (Gomez et al. 2012). In order to assess the levels of PPCPs in water bodies, several analytical methods, such as liquid chromatography–mass spectrometry positive-ion electrospray (LC/MS-ESI(+)), high-performance liquid chromatography (HPLC), and gas chromatography–mass spectrometry (GC-MS), have been used (Kolpin et al. 2002).

Among many classes of PPCP pollutants, chemical ultraviolet filters (UV-F) deserve particular attention. These compounds are often used as cosmetics ingredients and polymeric material photostabilization agents. UV filters can be divided into two basic categories: organic and inorganic. The most common inorganic sunscreen agents are zinc oxide and titanium dioxide. On the other hand, organic UV filters comprise a wide group of compounds like cinnamates, benzophenones, salicylates, camphor derivatives, p-aminobenzoic acid, and benzimidazoles. It was estimated that the annual production of these compounds reached several thousand tons (Buser et al. 2006). The use of UV filters has grown recently as a result of increasing concern about sunlight exposure causing skin cancer (Gasparro 2000). Due to the popularity of sunscreen agents, numerous reports have appeared on their occurrence in the environment. Sunscreen agents are present in the surface waters (Straub 2002; Poiger et al. 2004; Tarazona et al. 2010), swimming pool waters (Cuderman and Heath 2007; Santos et al. 2012), drinking water (Loraine and Pettigrove 2006; Diaz-Cruz et al. 2012), wastewater (Damiani et al. 2006; Li et al. 2007; Rodil et al. 2012), and sewage sludge (De la Cruz et al. 2012; Zuloaga et al. 2012; Barón et al. 2013). The presence of UV filters in the environment poses a particular danger, since they are generally resistant to degradation during water treatment process (Oppenheimer and Stephenson 2006; Liu et al. 2012; Gao et al. 2013; Ramos et al. 2015).

Due to the low allergy risk (Kimura and Katoh 1995; Schauder and Ippen 1997) and good UV absorption properties (Santos et al. 2012), EHMC is one of the most popular sunscreen agent, which undoubtedly have an impact on the environment. Straub et al. (2002) found that the average concentration of EHMC in the River Rhine was 5.5 ng/l. According to the same study, EHMC was found in the tissues of different fish species (Straub et al. 2002). Since most UV filters are lipophilic, they are able to bioaccumulate in the human body and aquatic organisms (Balmer et al. 2005; Fent et al. 2010). The negative effect of UV filter accumulation in living organisms was confirmed by many studies. Paredes et al. (2014) found that popular sunscreen agents, benzophenone-3 (BP-3), benzophenone-4 (BP-4), 2-ethylhexyl 4-methoxycinnamate (EHMC), and 4-methylbenzylidene-camphor (4-MBC), are significantly toxic for several different marine organisms. Furthermore, UV filters can cause endocrine disorders (Schlumpf et al. 2001; Heneweer et al. 2005; Kunz et al. 2006; Schlecht et al. 2006; Brausch et al. 2011). Sunscreens agent endocrine disrupting activity was described for human cells (Schlumpf et al. 2001; Heneweer et al. 2005), recombinant yeast with the human receptor of estrogen (Kunz et al. 2006), rodents (Schlumpf et al. 2001; Schlecht et al. 2006), and aquatic organisms (Brausch et al. 2011).

One of the routes by which UV filters and their chlorinated by-products may enter the environment is through disposal of the chlorinated water from swimming pools. According to the literature (Nakajima et al. 2009; Santos et al. 2013), the main by-products of EHMC transformations in the presence of HOCl are chlorosubstituted EHMC derivatives. These compounds are formed via electrophilic aromatic substitution reaction (Santos et al. 2013). Generally, there are two types of chlorination mechanisms occurring under water treatment conditions: electrophilic substitution and addition to unsaturated bond (Deborde et al. 2008). According to the Fukui function concept which is based on the frontier molecular orbital (FMO) theory and density functional theory (DFT), susceptibility of particular atoms to the nucleophilic, electrophilic, and radical attack can be directly evaluated from the atomic charge population analysis (Yang and Mortier 1986). Noteworthy, Fukui function has been applied in many important areas such as environmental fate assessment (zen et al. 2003; De Witte et al. 2009; Butler et al. 2010; Barr et al. 2012; Rokhina et al. 2012; Altarawneh et al. 2015), pollutant sorption and catalytic degradation (Chatterjee et al. 2000; Chatterjee et al. 2002; Chatterjee et al. 2003; Bekbolet et al. 2009; Musa et al. 2012; Pan et al. 2014; Palma-Goyes et al. 2015), and toxicity prediction of chlorinated benzenes (Padmanabhan et al. 2005, 2006), biphenyls (Parthasarathi et al. 2003; Parthasarathi et al. 2004; Padmanabhan et al. 2005), dibenzofuranes (Sarkar et al. 2006), and phenols (Padmanabhan et al. 2006).

Although it has been reported that EHMC undergoes degradation in the presence of UV light and different oxidizing and chlorinating agents (Nakajima et al. 2009; MacManus-Spencer et al. 2011; Gackowska et al. 2014), there is no comprehensive report dealing with the formation of EHMC chlorinated breakdown products under the water chlorination conditions. However, this is a complex issue, since there are many factors which can affect degradation of UV filters such as sunlight radiation, pH, and dissolved organic and inorganic compounds (Sakkas et al. 2003, Nakajima et al. 2009; Jammoul et al. 2009; Zhang et al. 2010; Zhou et al. 2013). The present first-principle study aims to determine what breakdown products are formed in case of the simple model containing EHMC and disinfecting agent. In order to gain better insight into this question, experimental results were supported with Fukui function value analysis.

## Materials and methods

### Materials

All chemicals were used without further purification. 2-Ethylhexyl 4-methoxycinnamate (EHMC) (98 %) and 4-methoxycinnamic acid (MCA) (99 %) were obtained from ACROS Organics and were kept in lightproof container at 4 °C. Other chemicals used in the study, sodium hypochlorite (100 g/l available Cl) anhydrous $Na_2SO_4$, dichloromethane, methanol (96 %) were purchased from POCH S.A. (Poland). Ethyl acetate and n-hexane were obtained from Sigma-Aldrich.

### Reaction conditions and analytical procedures

One thousand milliliters of EHMC/NaOCl and MCA/NaOCl reaction mixtures were prepared by dissolving EHMC and MCA in 1 ml of methanol, diluting with distilled water and adding NaOCl water solution. This procedure was based on the approaches reported in previous works (Giokas et al. 2004; Nakajima et al. 2009; MacManus-Spencer et al. 2011) according to which small amounts of methanol are used in order to improve EHMC solubility. The concentration of EHMC and MCA in the mixture was $3.4 \cdot 10^{-4}$ M, while the concentration of sodium hypochlorite was $1.7 \cdot 10^{-5}$ M. Immediately after the mixtures were prepared, they were transferred to the reactor. The temperature in the reactor was ranged from 22 to 25 °C. During the reaction, samples of 100 ml were collected at different time intervals and subjected to the extraction using 20 ml of extracting agent (n-hexane, ethyl acetate/n-hexane 1:1 mixture and dichloromethane). This procedure was carried out for 10 min. Then, the extracts were allowed to stand for 30 min at room temperature and dried using anhydrous $Na_2SO_4$. So, prepared samples were analyzed by GC-MS method. The temperature and pH of reaction mixtures were measured at each sample collection using a pH meter equipped with a temperature probe (CX-501 Multifunction Meter Elmetron, Poland). All reactions (with and without UV radiation exposure) were carried out in a stirred (200 rpm) and water-cooled cylindrical glass photoreactor of 650-ml capacity (Heraeus, Germany) equipped with a medium pressure mercury lamp (Heraeus, TQ 150 W), which can emit radiation in the 200–600-nm range (radiant flux 6.2 W for UV-C, 3.6 W for UV-B, and 4.5 W for UV-A). In case of photodegradation reactions, the samples were irradiated during all the experimental time (180 min.).

A GC-MS 5890 HEWLETT PACKARD instrument equipped with column ZB-5MS (0.25 mm × 30 m × 0.25 μm) was used for the identification of the transformation products applying the following chromatographic conditions: injector temperature 250 °C, oven temperature program from 80 °C to 260 °C at 10 °C/min, from 260 °C to 300 °C (held for 2 min) at a rate of 5 °C min. Helium was used as a carrier gas. The volume of the sample was 1 μl. Reaction products were identified by comparing recorded MS spectra with standard spectra from NIST/EPA/NIH Mass Spectral Library. This was done automatically using probability-based matching (PBM) mass spectrometry library search algorithm provided by HP ChemStation 3.0 software package.

### Fukui function calculations

All geometry optimizations were performed using Becke three-parameter (B3) hybrid method (Becke 1988; Becke 1993) with the Lee–Yang–Parr (LYP) functional (Lee et al. 1988; Miehlich et al. 1989) and 6-31+G(d,p) basis set (Krishnan et al. 1980; McLean and Chandler 1980; Clark et al. 1983; Frisch et al. 1984). Since the molecular geometry of substrates depends on the interactions between the molecule and surrounding solvent, polarized continuum model (PCM) was used (Miertuš et al. 1981; Miertuš and Tomasi 1982). Frequency calculations were performed on the B3LYP/6-31+G(d,p) level and no-imaginary ones were found. All geometry optimizations and frequency calculations were conducted using Gaussian03 software (Frisch et al. 2003).

Nucleophilic, $f^+$, electrophilic $f^-$, and radical $f^0$ Fukui function values were calculated within $DMOL^3$ (Delley 1990; Delley 1996; Delley 2000) module of Accelrys Material Studio 7.0 package (Accelrys Material Studio-Inc, 2014) using BLYP functional (Becke 1988; Lee et al. 1988; Miehlich et al. 1989) with DND basis set (version 3.5) (Delley 2006). Population analysis was performed using Hirshfeld method. The reliability of this procedure was confirmed by Thanikaivelan et al. (2002). Condensed nucleophilic, $f^+$, electrophilic $f^-$, and radical $f^0$ Fukui functions were calculated from Eq. (1–3) according to the method proposed by Yang and Mortier (Yang and Mortier 1986):

$$f^+ = q(N+1) - q(N) \qquad (1)$$
$$f^- = q(N) - q(N-1) \qquad (2)$$
$$f^0 = \frac{q(N+1) - q(N-1)}{2} \qquad (3)$$

where $q$ is Hirshfeld charge population on atoms in the molecule and $N$ denotes number of electrons in optimized structure. Numbering of atoms is presented on Fig. 1.

## Results and discussion

In order to evaluate the effect of HOCl on the stability and photostability of EHMC, GC-MS analyses of reaction mixtures were performed. Since MCA is known to be EHMC degradation product (MacManus-Spencer et al. 2011; Gackowska et al. 2014), reaction of MCA with HOCl was also examined. As it was mentioned, products were identified using PBM mass spectrometry library search method. For the analysis, only spectra with match probabilities above 95 % were taken into account. Retention times and selected $m/z$ peaks are summarized in Table 1. MS spectra of identified products are provided in the Online Resource.

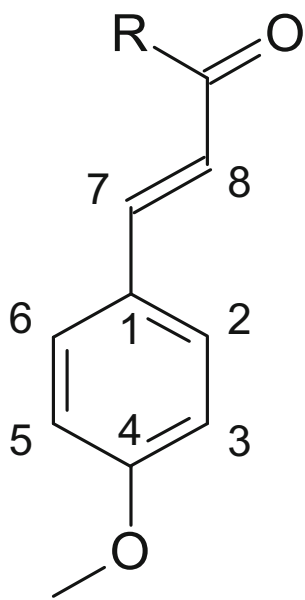

**Fig. 1** Atom numbering in MCA (R=OH) and EHMC (R=OCH$_2$CH(C$_2$H$_5$)C$_4$H$_9$-n)

Since the pH of obtained mixtures was ranged from 7.1 to 7.4 in case of EHMC/NaOCl and from 6.9 to 7.2 in case of MCA/NaOCl, according to Henderson–Hasselbalch equation, HOCl is a dominating form (p$K_a$=7.5). As it might be expected, under these conditions, EHMC undergoes chlorination which was also reported previously (Nakajima et al. 2009; Santos et al. 2013). GC-MS analysis showed that there are two compounds of molecular mass of 324 Da (1EHMCCl and 2EHMCCl) (Table 1), which probably correspond to chlorinated EHMC derivatives. Noteworthy, $m/z$ signals of 1EHMCCl and 2EHMCCl (Fig. S4 and Fig. S5 in the Online Resource) are in accordance with monochlorinated EHMC fragmentation pattern reported in the literature (Nakajima et al. 2009; Santos et al. 2013). Low intense molecular peaks and the loss of 112 $m/z$ units on the mass spectra of EHMC and its monochlorinated derivatives (supplementary Fig. S1, Fig. S2, Fig. S4, Fig. S5) indicates fast fragmentation wherein the first step is the McLafferty rearrangement. As one might expect, MS spectra recorded for E-EHMC and Z-EHMC are almost identical, and hence, they can be distinguished only by their retention times (Table 1). Z-EHMC, which is known to be more polar than E isomer (Pattanaargson et al., 2004), is represented on the GC chromatogram by small peak of a lower retention time that E-EHMC (Fig. 2a). As in case of EHMC and Z-EHMC, 1EHMCCl and 2EHMCCl have very similar mass spectra but different retention times indicating they are also geometric isomers. According to Nakajima et al. 2009, peaks at $m/z$ 195 and 212 in the monochlorinated EHMC MS spectra correspond to [C$_{10}$H$_8$ClO$_2^+$] and [C$_{10}$H$_9$ClO$_3^+$] fragments (Nakajima et al. 2009). Since there are two of the most abundant chlorine isotopes $^{35}$Cl and $^{37}$Cl, 195 and 212 $m/z$ peaks on the 1EHMCCl and 2EHMCCl mass spectra are near to the less intense peaks having $m/z$ of 197, and 214 respectively (supplementary Fig. S4 and Fig S5). This chlorine isotope signature is typical for monochlorinated compounds. Therefore, as one can see from Fig. S6 in the Online Resource, two molecular peaks at $m/z$=142 and 144 can be found on the 1-chloro-4-methoxybenzene MS spectra. On the other hand, in case of dichlorinated molecules due to the three combinations of chlorine isotopes ($^{35}$Cl/$^{35}$Cl, $^{35}$Cl/$^{37}$Cl, and $^{37}$Cl/$^{37}$Cl), there are three MS signals comprising the chlorine isotope profile. This pattern can be observed for molecular peaks of 1,3-dichloro-2-methoxybenzene ($m/z$=176, 178, and 180), 2,4-dichlorophenole ($m/z$=162, 164, and 166), and 2,6-dichloro-1,4-benzoquinone ($m/z$=176, 178, and 180) (supplemental Fig. S7, Fig. S9, and Fig. S10).

As it can be inferred from Table 1, the diversity of detected products is dependent on the solvent used in the extraction step. In order to examine this question, the extraction procedure was performed using non-polar solvent, n-hexane, and polar solvents, dichloromethane and ethyl acetate/n-hexane (1:1) It is worth mentioning that acetate/n-hexane system was found to be a good extracting agent for mixtures containing polar and non-polar compounds (Strati and Oreopoulou 2011). Furthermore, binary solvents such as ethyl acetate/n-hexane and dichloromethane/ethyl acetate were applied in case of EHMC and octyl dimethyl-p-aminobenzoate (ODPABA) chlorinated products extraction (Sakkas et al. 2003; Nakajima et al. 2009). The choice of a suitable solvent is of course related to the polarity of analytes. According to our results, more compounds were detected when ethyl acetate/n-hexane and dichloromethane were used than when the samples were extracted with non-polar n-hexane (Table 1). Noteworthy, good extracting properties of dichloromethane were confirmed in case of both hydrophilic compounds like chlorophenols (Yasman et al. 2006) and hydrophobic compounds like fatty acids and sterols (Chen et al. 1981). On the basis of retention time analysis, MacManus-Spencer et al. (2011) found that most of degradation products formed via EHMC photolysis are generally more polar than EHMC. A similar observation can be made in case of EHMC/HOCl system, since GC-MS retention time values of chlorinated degradation products like 1-chloro-4-methoxybenzene, 1,3-dichloro-2-methoxybenzene, 2-ethylhexyl chloroacetate, and 3-chloro-4-methoxybenzaldehyde are lower than retention times of E-EHMC and Z-EHMC (Table 1).

The effect of UV radiation on the stability of the EHMC has been described in many studies (Morlière et al. 1982; Pattanaargson and Limphong 2001; Pattanaargson et al.,

**Table 1** GC retention times and selected mass spectra data of detected in the reaction mixtures by-products

| Compound | Retention time [min] | Proposed formulae | Calculated logP[a] | Selected m/z signals (% of base peak) |
|---|---|---|---|---|
| EHMC/NaOCl (n-hexane extract) | | | | |
| Z-EHMC | 18.32 | $C_{18}H_{26}O_3$ | 5.38 | 290 (6), 178 (100), 161 (62) |
| E-EHMC | 19.59 | $C_{18}H_{26}O_3$ | 5.38 | 290 (6), 178 (100), 161 (60) |
| EHMC/NaOCl (ethyl acetate/n-hexsane extract) | | | | |
| 2-Ethylhexyl chloroacetate | 8.74 | $C_{10}H_{19}ClO_2$ | 3.48 | 112 (15), 83 (29), 70 (84), 57 (100) |
| Z-EHMC | 18.22 | $C_{18}H_{26}O_3$ | 5.38 | 290 (5), 178 (100), 161 (59) |
| 1EHMCCl | 19.22 | $ClC_{18}H_{25}O_3$ | 5.98 | 324 (17), 212 (100), 176 (63) |
| E-EHMC | 19.89 | $C_{18}H_{26}O_3$ | 5.38 | 290 (6), 178 (100), 161 (61) |
| 2EHMCCl | 21.31 | $ClC_{18}H_{25}O_3$ | 5.98 | 324 (16), 212 (100), 176 (73) |
| EHMC/NaOCl (dichloromethane extract) | | | | |
| 1-Chloro-4-methoxybenzene | 5.54 | $ClC_6H_4OCH_3$ | 2.42 | 142 (100), 127 (71), 99 (98) |
| 1,3-Dichloro-2-methoxybenzene | 8.14 | $Cl_2C_6H_3OCH_3$ | 3.02 | 176 (95), 161 (100), 133 (98) |
| 2-Ethylhexyl chloroacetate | 8.85 | $C_{10}H_{19}ClO_2$ | 3.48 | 112 (17), 83 (31), 70 (86), 57 (100) |
| 3-Chloro-4-methoxybenzaldehyde | 10.70 | $ClC_6H_3(CHO)OCH_3$ | 2.13 | 169 (100), 141 (13), 126 (21) |
| Z-EHMC | 18.34 | $C_{18}H_{26}O_3$ | 5.38 | 290 (6), 178 (100), 161 (60) |
| 1EHMCCl | 19.36 | $ClC_{18}H_{25}O_3$ | 5.98 | 324 (17), 212 (100), 176 (61) |
| E-EHMC | 20.0.3 | $C_{18}H_{26}O_3$ | 5.38 | 290 (6), 178 (100), 161 (61) |
| 2EHMCCl | 21.45 | $ClC_{18}H_{25}O_3$ | 5.98 | 324 (11), 212 (100), 176 (74) |
| MCA/NaOCl (ethyl acetate/n-heksane extract) | | | | |
| 2,4-Dichlorophenole | 6.32 | $Cl_2C_6H_3OH$ | 2.88 | 162 (100), 126 (18), 98 (50) |
| 2,6-Dichloro-1,4-benzoquinone | 7.12 | $Cl_2C_6H_2O_2$ | 1.80 | 176 (77), 148 (18), 120 (52), 88 (86), 60 (100), 53 (90) |
| 1,3-Dichloro-2-methoxybenzene | 7.88 | $Cl_2C_6H_3OCH_3$ | 3.02 | 176 (98), 161 (100) 133 (97) |
| 1,2,4-Trichloro-3-methoxybenzene | 8.46 | $Cl_3C_6H_2OCH_3$ | 3.63 | 210 (62), 195 (100), 167 (70) |
| 2,4,6-Trichlorophenole | 8.89 | $Cl_3C_6H_2OH$ | 3.48 | 196 (100), 160 (18) 132 (59) |
| 3,5-Dichloro-2-hydroxyacetophenone | 15.21 | $Cl_2C_6H_2(OH)COCH_3$ | 3.09 | 189 (100), 123 (41), 75 (46) |
| EHMC/UV (ethyl acetate/n-heksane extract) | | | | |
| 2-Ethylhexyl alcohol | 4.28 | $C_8H_{18}O$ | 2.50 | 112 (5), 98 (7), 83 (17), 70 (22), 57 (100) |
| Z-EHMC | 18.29 | $C_{18}H_{26}O_3$ | 5.38 | 290 (6), 178 (100), 161 (63) |
| E-EHMC | 19.88 | $C_{18}H_{26}O_3$ | 5.38 | 290 (7), 178 (100), 161 (60) |
| EHMC/NaOCl/UV (ethyl acetate/n-heksane extract) | | | | |
| 2-Ethylhexyl alcohol | 4.22 | $C_8H_{18}O$ | 2.50 | 112 (6), 98 (7), 83 (18), 70 (25), 57 (100) |
| Z-EHMC | 18.20 | $C_{18}H_{26}O_3$ | 5.38 | 290 (7), 178 (100), 161 (61) |
| 1EHMCCl | 19.11 | $ClC_{18}H_{25}O_3$ | 5.98 | 324 (18), 212 (100), 176 (63) |
| E-EHMC | 19.73 | $C_{18}H_{26}O_3$ | 5.38 | 290 (6), 178 (100), 161 (62) |
| 2EHMCCl | 21.15 | $ClC_{18}H_{25}O_3$ | 5.98 | 324 (10), 212 (100), 176 (76) |

[a] logP values were calculated using Marvin Sketch (https://www.chemaxon.com/)

2004; Rodil et al. 2009; Mac-Manus-Spencer et al. 2011; Miranda et al. 2014). According to our results, it takes ca. 60 min until the E–Z equilibrium is reached (Fig. 3a) under chlorine-free UV treatment conditions. In case of, EHMC/HOCl/UV (Fig. 3b) at the beginning of the process, E–Z isomerisation is the main reaction and after the E–Z equilibrium is reached, Z-EHMC undergoes transformations leading to the monochlorosubstituted derivatives and photodegradation products (Table 1). Surprisingly, much less breakdown products are formed when the EHMC/HOCl mixture was subjected to the UV light then when the reaction was carried out in the dark (Fig. 2). As one can see, the intensities of GC peaks of EHMC/HOCl are generally higher than appropriate signals recorded for EHMC/HOCl/UV. This shows that EHMC degradation is more effective when UV irradiation is applied. According to many studies, exposure to the UV light combined with $Cl_2$, $O_3$, and $H_2O_2$ addition generally gives better results in terms of the removal of pollutants than in cases when UV irradiation was not used (Wols et al. 2013; Gong et al. 2015; Nam et al. 2015).

**Fig. 2** Exemplary gas chromatograms of EHMC/HOCl (**a**) and EHMC/HOCl/UV (**b**)

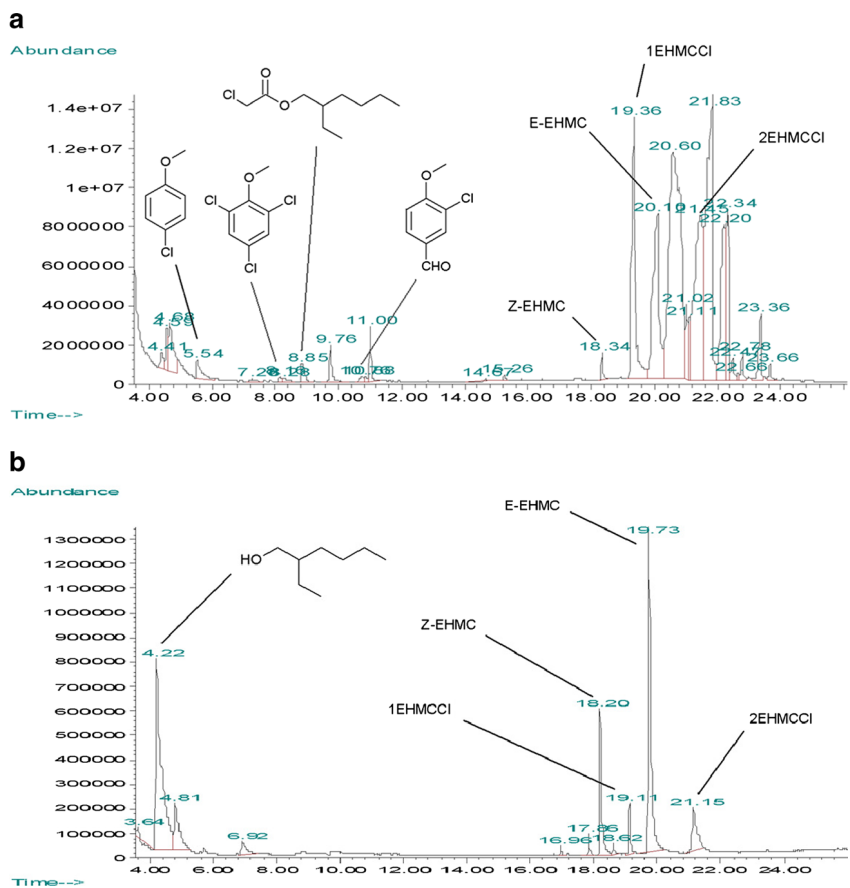

According to prior studies, EHMC can undergo photolysis leading to MCA and 2-ethylhexyl alcohol (Nakajima et al. 2009; Manus-Spencer et al. 2011; Gackowska et al., 2014). Obviously, MCA and 2-ethylhexyl alcohol can be formed also through the hydrolysis when the reaction mixture is not exposed to the UV radiation. Interestingly, reaction of MCA with HOCl results in the formation of acetophenone derivative, 3,5-dichloro-2-hydroxyacetophenone. According to Hilton et al. (1990), acetophenone can be formed via cinnamic acid degradation according to the mechanism which proceeds in several steps, including water addition to C=C bridge, dehydrogenating oxidation, tautomerization, and decarboxylation. The high reactivity of C=C bond in cinnamic acid derivatives can be observed also in case of EHMC. As it can be seen from Fig. 4, although in case of EHMC, a large COOCH2CH(C2H5)C4H9-n group is attached to C=C bond, there is no steric effect that would hinder electrophilic attack to C=C bridge. Besides, negligible steric effect can be also confirmed by the fact that EHMC undergoes [2+2] cycloaddition reaction and radical attack of $^1O_2$ resulting in the formation of 4-methoxybenzaldehyde and 2-ethylhexyl acetate (MacManus-Spencer et al. 2011).

The abovementioned examples suggest that C=C bridge is highly susceptible for electrophilic and radical attack. This can be explained using Fukui function analysis (Table 2) performed for optimized E-EHMC, Z-EHMC, E-MCA, and Z-MCA structures (Fig. 4). The greater the Fukui function value of a given atom is, the higher its local reactivity. In case of EHMC and MCA molecules, electron donating (OMe) group is conjugated with electron withdrawing carboxylic and ester group promoting electrophilic attack to C-3 and C-5 carbon atoms (atoms numbering according to Fig. 1). As one can see (Table 2), electrophilic Fukui function $f^-$ values calculated for C-3 and C-5 atoms are higher than those calculated for C-2 and C-6 atoms which is in accordance with electron donating/withdrawing nature of attached substituents. Fukui function analysis showed that C-8 atom is even more susceptible for electrophilic attack than C-3 and C-5 carbon atoms. Interestingly, C-8 atom is also the most reactive in terms of radical attack. Although HOCl exhibits mainly electrophilic properties (Deborde et al. 2008), OH$^\bullet$ and Cl$^\bullet$ can be formed during HOCl photolysis (Feng et al. 2007). Nevertheless, we did not detect C7–C8 bond cleavage products in case of EHMC/HOCl/UV system.

The accumulation of negative charge on the C-8 atom as compared to phenyl ring can be explained using resonance theory. According to one of the "resonance theory rules," the larger the distance between separated formal charges, the more energetically preferable structure and in turns its contribution to the resonance hybrid is higher. As it can be seen

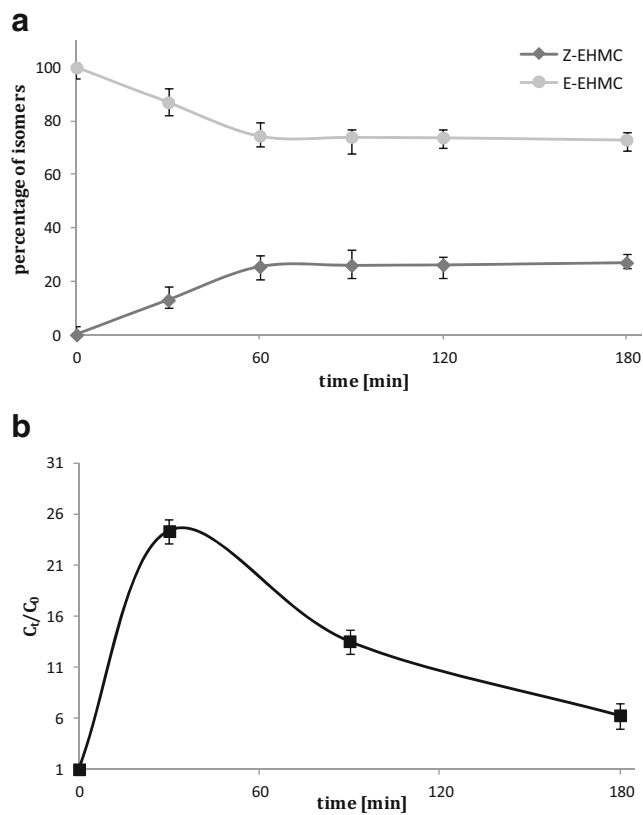

**Fig. 3** Photoisomerisation of EHMC (**a**) and degradation of Z-EHMC in the presence HOCl and UV irradiation (**b**)

from Fig. 5, there is better charge separation in case of structure 3 than in case of structure 2.

The products detected in the EHMC/HOCl reaction mixtures suggest that EHMC degradation probably proceeds in many steps including chlorination, oxidation, and decarboxylation (Fig. 6). Although 3,5-dichloro-4-methoxybenzaldehyde and 3,5-dichloro-4-methoxybenzoic acid were not detected in the reaction mixture, formation of 1,3-dichloro-2-methoxybenzene can be explained by the mechanism in which formyl group oxidation is followed by the decarboxylation. Noteworthy, the loss of carboxylic group attached to benzene ring can occur during water chlorination, since it was observed in case of 4-hydroxybenzoic acid (Larson and Rockwell 1979). It seems to be probable that some compounds which were formed in case of MCA/HOCl can be also formed in case of EHMC/HOCl. The products identified during MCA degradation suggest that 1,3-dichloro-2-methoxybenzene (which was also detected in EHMC/HOCl reaction mixture) can be further chlorinated to 1,2,4-trichloro-3-methoxybenzene (Table 1). The presence of chlorophenols and 2,6-dichloro-1,4-benzoquinone in the MCA/HOCl reaction mixture can be explained by the fact that anisole derivatives, formed during EHMC and MCA degradation, can be demethylated under the water chlorination conditions (Lebedev et al. 2004) and then oxidized to quinones (Pi and Wang 2007).

## Conclusions

In this study, new insight into the formation of EHMC disinfection by-products was presented. Reaction of EHMC with HOCl results in the formation of C=C bridge cleavage products, 2-ethylhexyl chloroacetate, 1-chloro-4-methoxybenzene, 1,3-dichloro-2-methoxybenzene, and 3-chloro-4-methoxybenzaldehyde. Interestingly, 2-ethylhexyl chloroacetate was not detected in the EHMC/HOCl/UV mixture. In order to give a complete view of EHMC transformation under the influence of HOCl the reaction of EHMC hydrolysis and photolysis product, MCA (MacManus-Spencer et al. 2011; Gackowska et al. 2014) with HOCl was examined.

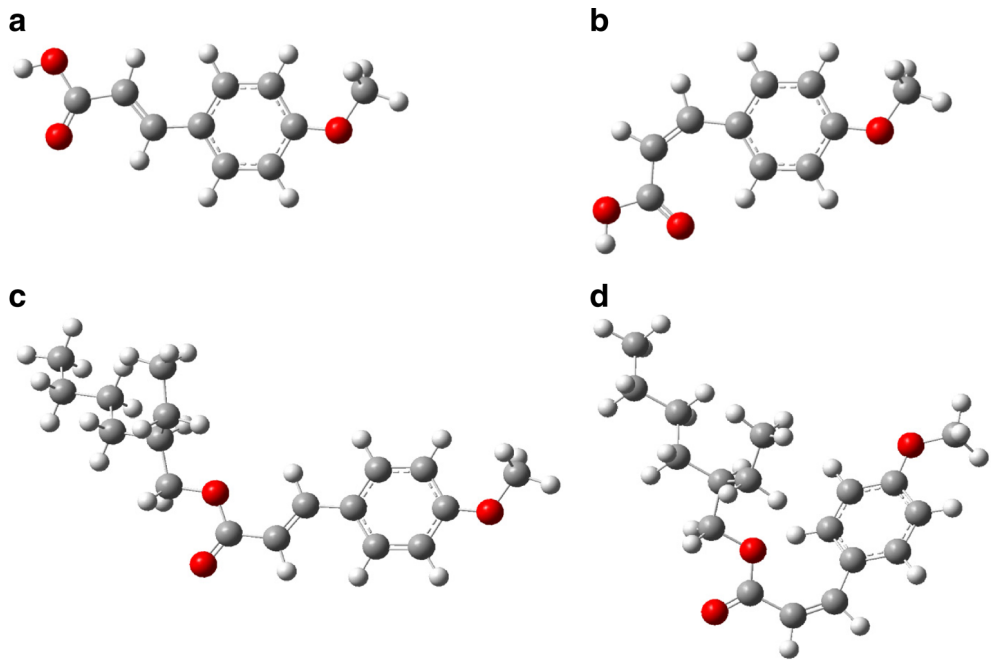

**Fig. 4** Visual representation of optimized E-MCA (**a**), Z-MCA (**b**), E-EHMC (**c**), and Z-EHMC (**d**) molecular structures

**Table 2** Nucleophilic, $f^+$, electrophilic $f^-$, and radical $f^0$ Fukui function values calculated for E-MCA, Z-MCA, E-EHMC, and Z-EHMC

| Compound | Atom | $f^+$ | $f^-$ | $f^0$ |
|---|---|---|---|---|
| E-MCA | C-1 | 0.054 | 0.047 | 0.050 |
|  | C-2 | 0.024 | 0.043 | 0.047 |
|  | C-3 | 0.041 | 0.069 | 0.042 |
|  | C-4 | 0.041 | 0.062 | 0.052 |
|  | C-5 | 0.062 | 0.062 | 0.062 |
|  | C-6 | 0.037 | 0.051 | 0.044 |
|  | C-7 | 0.105 | 0.037 | 0.071 |
|  | C-8 | 0.084 | 0.102 | 0.093 |
| Z-MCA | C-1 | 0.020 | 0.071 | 0.045 |
|  | C-2 | 0.039 | 0.037 | 0.038 |
|  | C-3 | 0.037 | 0.059 | 0.048 |
|  | C-4 | 0.061 | 0.061 | 0.061 |
|  | C-5 | 0.038 | 0.057 | 0.047 |
|  | C-6 | 0.051 | 0.051 | 0.051 |
|  | C-7 | 0.109 | 0.036 | 0.072 |
|  | C-8 | 0.083 | 0.105 | 0.094 |
| E-EHMC | C-1 | 0.025 | 0.067 | 0.046 |
|  | C-2 | 0.044 | 0.038 | 0.041 |
|  | C-3 | 0.037 | 0.053 | 0.045 |
|  | C-4 | 0.061 | 0.061 | 0.061 |
|  | C-5 | 0.037 | 0.055 | 0.046 |
|  | C-6 | 0.048 | 0.050 | 0.049 |
|  | C-7 | 0.101 | 0.039 | 0.070 |
|  | C-8 | 0.086 | 0.104 | 0.095 |
| Z-EHMC | C-1 | 0.019 | 0.072 | 0.046 |
|  | C-2 | 0.035 | 0.039 | 0.037 |
|  | C-3 | 0.034 | 0.056 | 0.045 |
|  | C-4 | 0.056 | 0.062 | 0.059 |
|  | C-5 | 0.035 | 0.056 | 0.045 |
|  | C-6 | 0.045 | 0.051 | 0.048 |
|  | C-7 | 0.109 | 0.035 | 0.072 |
|  | C-8 | 0.088 | 0.096 | 0.092 |

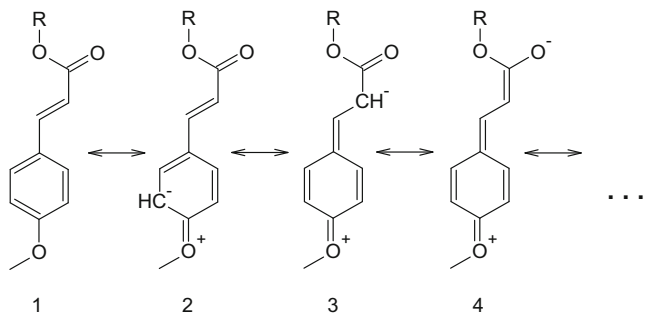

**Fig. 5** Selected resonance structures of MCA (R=H) and EHMC (R=CH$_2$CH(C$_2$H$_5$)C$_4$H$_9$-n)

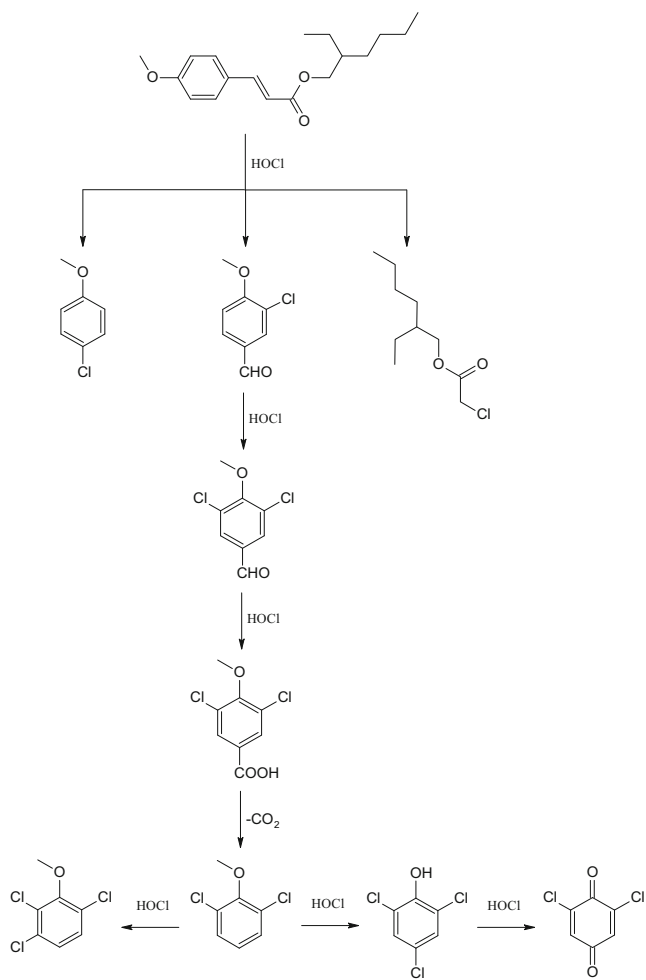

**Fig. 6** Proposed pathways of EHMC chlorinated breakdown products formation based on products detected in EHMC/HOCl and MCA/HOCl reaction mixtures

According to local reactivity analysis based on the Fukui function calculations, C=C bond attached to the benzene ring in EHMC and MCA molecules is highly susceptible for electrophilic and radical attack. As it was reported previously, specific reactivity of C=C bridge in case of cinnamic acid and its derivatives is manifested by [2+2] cycloaddition and photodegradation reactions (Egerton et al. 1981; Lewis et al. 1988; Robinet et al. 1987; Schrader et al. 1994; Broadbent et al. 1996; Hauri et al. 2004; MacManus-Spencer et al. 2011) and water addition (Hilton et al. 1990). Although it cannot be excluded that in case of EHMC/HOCl system, formation of C=C bond cleavage products is radical in nature, electrophilic attack of HOCl seems to be the initial step of degradation mechanism.

GC/MS measurements showed that EHMC disinfection by-products are more polar than EHMC. For these reason, much more products were detected when dichloromethane and ethyl acetate/n-hexane 1:1 mixture was used as extracting solvent than n-hexane. The polarity of compounds is a relevant parameter for both analytical procedure optimization and environmental risk assessment. Although hydrophilic chlorinated compounds not easily diffuse through cell membranes as hydrophobic compounds, they are also dangerous for water ecosystems. Chlorophenols and 2,6-dichloro-1,4-benzoquinone found in the MCA/HOCl reaction mixture are well-known toxic disinfection by-products which were identified in drinking water (Turnes et al. 1996; Michałowicz 2005; Simões et al. 2007; Qin et al. 2010). On the other hand, according to our best knowledge, there are no environmental monitoring reports on the 2-ethylhexyl chloroacetate occurrence in water bodies. This chlorinated EHMC breakdown product can be hydrolysed to chloroacetic acid which is known to be fitotoxic and cause serious environmental problems (Laturnus et al. 2005; Frank et al. 1994). Generally speaking, studies on the degradation of organic compounds in the presence of HOCl are useful for assessment of potential chloroorganic pollutant sources. Presented in this paper, results provide a good starting point for further studies on the UV filters chlorinated breakdown products formation.

**Acknowledgments** The authors gratefully acknowledge the help of Academic Computer Centre in Gdańsk in providing its facility to perform calculations presented in this study.